\documentclass[a4paper,oneside,11pt,draft]{article}
\usepackage{amsmath,amssymb,sectsty,mathrsfs}
\newcommand{\nc}{\newcommand}
\newcounter{appendixc}
\renewcommand{\appendix}[1]{\par\vspace{12pt}
\refstepcounter{appendixc}
\setcounter{equation}{0}
\renewcommand{\theequation}{\Alph{appendixc}.\arabic{equation}}
 \noindent
  \begin{center}
{\bf  Appendix \Alph{appendixc}: \ #1}
 \par \vspace{6pt}  
  \end{center}}
\nc{\cc}{\mathrm{c.c.}}\nc{\be}{\begin{equation}}
\renewcommand{\Re}{\mathrm{Re}\,}

\nc{\beg}{\begin{equation*}}\nc{\eeg}{\end{equation*}}
\nc{\ee}{\end{equation}}
\nc{\frec}[2]{{\textstyle{\frac{#1}{#2}}}}
\nc{\ud}{\mathrm{d}}\nc{\Tr}{\mathrm{Tr}\,}
\nc{\heq}{\,\,\hat{=}\,\,}\nc{\nheq}{\,\,\hat{\neq}\,\,}
\nc{\const}{\mathrm{const}}\nc{\bxi}{\bar{\xi}}
\nc{\bm}{\bar{m}}\nc{\bz}{\bar{z}}\nc{\bzeta}{\bar{\zeta}}
\nc{\nn}{\tilde{n}}\nc{\nt}{\tilde{t}}\nc{\hM}{\hat{M}}
\nc{\bF}{\bar{F}}\nc{\hP}{\hat{P}}\nc{\hu}{\hat{u}}
\nc{\sgn}{\mathrm{sgn}}\nc{\cov}{{\scriptscriptstyle |}}
\nc{\scri}{\mathscr{I}^+}\nc{\norma}{|\!|}
\nc{\contr}{-\!\!\!\lower-.260em\hbox{$\lrcorner$}\,}\nc{\bzet}{\bar{z}}
\nc{\bes}{\begin{subequations}}\nc{\ees}{\end{subequations}}
\nc{\mcr}{ R }\nc{\vol}[1]{\,\text{vol}(#1)\,}
\nc{\mch}{\mathcal{H}}\nc{\mcm}{\mathcal{M}}
\nc{\mcs}{\mathcal{S}}
\allsectionsfont{\normalsize}
\newtheorem{tw}{Proposition}[section]
\begin{document}

\title{Horizons in Robinson-Trautman space-times}
\author{{\small W. Natorf, J. Tafel}\\
{\small Institute of Theoretical Physics, Warsaw University,}\\
{\small ul. Ho\.za 69, 00-681 Warsaw, Poland}\\
{\small e-mail: nat@fuw.edu.pl, tafel@fuw.edu.pl}}
\date{}
\maketitle
\begin{abstract}
The past quasi-local horizons in vacuum Robinson-Trautman
spa\-ce-ti\-mes are described.  The case of a null (non-ex\-pan\-ding)
horizon is discussed. It is shown that the only Robinson-Trautman
space-time admitting such a horizon with sections diffeomorphic to
$S_2$ is the Schwarzschild space-time. Weakening this condition leads
to the horizons of the C-metric. Properties of the hypersurface $r=2m$
are examined.
\end{abstract}
\section{Introduction}
A vacuum Robinson-Trautman (RT) metric can be determined by a function
of three variables, satisfying a fourth order partial differential
equation \cite{RT}. Very few of its solutions are known in an explicit
form \cite{KR}. The only ones representing asymptotically flat
space-times correspond to the Minkowski and the Schwarz\-schild
metrics. Nontrivial asymptotically flat RT space-times were shown to
exist and tend to the Schwarzschild solution in the limit of infinite
retarded time \cite{CHR}.

In this paper we describe quasi-local horizons \cite{AK} in RT
space-times. We derive equations obtained previously by Tod \cite{TOD}
and Chow and Lun \cite{CL}. Using the equation of Chow and Lun we
prove that the past horizon is null if and only if its sections admit
a shear-free inward pointing null normal vector field.

It is a well known property of a nonexpanding horizon that its null
generator is the repeated principal null direction of the Weyl tensor
\cite{AK}. Since the repeated principal null direction responsible for
algebraic properties of the Weyl tensor of RT space-time is transverse
to the horizon, the Weyl tensor is of Petrov type D on the horizon. We
prove that this property holds also in the neighbourhood of the
horizon and no nonexpanding horizons of the form $R\times S_2$ exist
unless the RT space-time is the Schwarzschild space-time.  If no
restriction on topology of sections is made, one obtains null horizons
of the C-metric. Their sections contain conical singularities
\cite{GKP}.

In section \ref{eh} we consider the metric induced on the hypersurface
$r=2m$ for finite $u$, and examine its signature (Proposition
\ref{rtnull}).  We also find the relation between this hypersurface
and trapped surfaces (Proposition \ref{r2m}).

\bigskip

\noindent The standard form of the Robinson-Trautman metric reads \cite{RT,KR}
\be\label{RTmetric} g = 2\ud u(H \ud u + \ud r) - 
2 \frac{r^2}{P^2}\ud\xi\ud\bar{\xi}\ .\ee
Vacuum Einstein's equations give 
\be\label{Heq}
H = P^2(\ln P)_{,\xi\bar{\xi}}-r(\ln P)_{,u}-\frac{m}{r}
\ee
and the Robinson-Trautman equation
\be
\label{RTeq}
P^2(P^2(\ln P)_{,\xi\bar{\xi}})_{,\xi\bar{\xi}}+3m (\ln P)_{,u}-m_{,u}=0\ .
\ee 
Here $m$ is a function of $u$ which, when non-vanishing,
 can be transformed to $m=\pm 1$.  In this case equation 
\eqref{RTeq} takes the form
\be
K_{,\xi\bar{\xi}}-3m(P^{-2})_{,u}=0\label{RT1}
\ee
where 
\be\label{K} K=2P^2(\ln P)_{,\xi\bar{\xi}}\ee 
is the Gaussian curvature of the surface given by  
$u=\const,\ r=1$.
Point transformations preserving metric
\eqref{RTmetric} and  $m$ have the
following form \cite{KR}:
\begin{equation}\begin{aligned}
\label{ttr}
&u  \rightarrow  u + \const \ ,\\
&\xi \rightarrow h(\xi) \ ,\\
&P  \rightarrow  |h'|^{-1} P  \ ,\\
& r  \rightarrow  r  \ .
\end{aligned}\end{equation}

The induced metric of  a surface of constant $u$ and $r$
is given by
\be\label{ru}
g' = 2\frac{r^2}{P^2}\ud\xi\ud\bar{\xi}\ .
\ee
If the
function
\be\label{phat}
\hP = P/(1+\xi\bar{\xi}/2)
\ee
is regular (smooth and positive) for all $\xi\in C$ and $\xi\to\infty$,
then the surface $u=\const,\ r=\const$ is diffeomorphic to $S_2$
and $\xi$ can be interpreted as the complex stereographic
coordinate on $S_2$. The function  $P = (1+\xi\bar{\xi}/2)$ giving the
standard metric of $S_2$ will be denoted by $P_S$. Two dimensional
surfaces admitting regular $\hP$ will be referred to as regular
surfaces.

\section{Marginally trapped surfaces}
Let $\mcs$ be a 2-dimensional space-like surface given by
\be\label{ciecie}
u = \const,\  r = \mcr(\xi,\bar{\xi})\ .\ee
We introduce four null forms, 
\begin{equation}\begin{aligned}
&\theta^0 = \ud u\ ,\\ 
\label{tetrad} &
\theta^1 = \left(H + \frac{P^2}{r^2}|\mcr_{,\xi}|^2\right)\ud u +
\ud r - (\mcr_{,\xi}\ud\xi+\cc)\ ,\\
& \theta^2 = \frac{r}{P}\ud\xi-\frac{P}{r}\mcr_{,\bar{\xi}}\ud u\ ,\quad 
\theta^3 = \overline{\theta^2}\ ,
\end{aligned}\end{equation}
and the dual null tetrad, 
\begin{equation}\begin{aligned}
& e_0 = \partial_u - \left(H-\frac{P^2}{r^2}|\mcr_{,\xi}|^2\right)\partial_r
+\frac{P^2}{r^2}(\mcr_{,\bar{\xi}}\partial_{\xi}+\cc)\ ,\\
&e_1 = \partial_r\ \label{dstetrad},\\
& e_2 = \frac{P}{r}(\partial_{\xi}+\mcr_{,\xi}\partial_r)\ ,\quad 
e_3 = \overline{e_2}\ .\end{aligned}
\end{equation}
In Appendix A we list the non-vanishing Newman-Penrose spin
coefficients and Weyl scalars related to the tetrad \eqref{dstetrad},
needed in further calculations.  The ingoing and outgoing null vectors
normal to $\mcs$ are, respectively, $l = e_0,\ k = e_1$, while $m=e_2$
and $\bar{m} = e_3$ are tangent to $\mcs$.

Using the formulae 
\begin{equation}\begin{aligned}
\label{exprozsz}
 -m^a\bar{m}^b l_{(a;b)} &= \frac{H}{r}+(\ln P)_{,u}-
\frac{P^2}{r^2}\mcr_{,\xi\bar{\xi}}+\frac{P^2}{r^3}|\mcr_{,\xi}|^2\ ,\\
 -m^a\bar{m}^b k_{(a;b)} & = -r^{-1}
\end{aligned}\end{equation}
from Appendix A and substituting for $H$ from \eqref{Heq}
we find the  expansion scalars of the vectors $k$ and $l$ on $\mcs$ \cite{TOD}
\be\label{expl}
\theta_{(l)} = \frac{1}{\mcr}\left(
\frac{K}{2}- P^2(\ln\mcr)_{,\xi\bar{\xi}}-\frac{m}{\mcr}\right)\ ,\ee
\be\label{expk}
\theta_{(k)}= -\mcr^{-1} \ .\ee
Therefore, the condition that the expansion of $l$ vanishes
takes the form of the following equation for $\mcr$:
\be\label{mts}
-P^2(\ln\mcr)_{,\xi\bar{\xi}}+\frac{K}{2} -\frac{m}{\mcr}=0\ .\ee
The existence and uniqueness of positive, smooth solutions of
\eqref{mts} for regular spherical surfaces $u=\const,\ r=\const$ has
been proved by Tod \cite{TOD}.  If $m=1$, the area of these surfaces
is independent of $u$ due to \eqref{RTeq} and Stokes' theorem, and can
be chosen to be equal to $4\pi$.  Note that $u$ enters \eqref{mts}
as a parameter via the function $P$.

\section{Horizons}
In this section we consider hypersurfaces foliated by marginally
trapped surfaces. Let $\mch$ be defined by
\be\label{horizon}
r = \mcr(u,\xi,\bar{\xi})\ ,\ee
where $R$ satisfies \eqref{mts}. It follows that
\be\label{evol}
\nabla_{t}(m^a \bar{m}^b l_{(a|b)}) =0\ee
on $\mch$, where $t$ is tangent to $\mch$ and orthogonal to its
sections.  Using maximum principle and \eqref{expl} Tod proved that a
marginally trapped surface described by \eqref{ciecie} and \eqref{mts}
is outermost \cite{TOD}.  Results of Andersson et al. \cite{AMS}
assure that there exists a horizon such that $\mcs$ is its section.

Using \eqref{dstetrad} we obtain
\be\label{tdec} t = l  + \chi k\ ,\quad
\chi = \mcr_{,u}+H+\frac{P^2}{\mcr^2}
\mcr_{,\xi}\mcr_{,\bar{\xi}}\ .\ee
Following \cite{CL} we rewrite equation \eqref{evol} in terms of the spin
coefficients:
\be\label{evolnp}
\frac{P^2}{\mcr}\left(\frac{\chi}{\mcr}\right)_{,\xi\bar{\xi}}+
\chi\Psi_2 - \lambda\bar{\lambda} = 0\ ,
\ee
where $\lambda$ is the shear of $l$. Equation \eqref{evolnp} corresponds
to equation (15) in \cite{MK}, where the general case of constraint
equations on a marginally trapped tube was considered.

Assume now that the surfaces $u=\const,\ r=\const$ have spherical
topology. As pointed out in \cite{CL}, it follows from the maximum
principle for \eqref{evolnp} that $\mch$ is a non-timelike surface.
If $\mch$ is null ($\chi =0 $), then $\lambda=0$. Conversely, if
$\lambda$ vanishes, multiplying both sides of \eqref{evolnp} by
$\chi/\mcr^2$ and using Stokes' theorem for $\mcs$ we get
\be\label{contr}
-\int_{\mcs}\left|\nabla\frac{\chi}{\mcr}\right|^2 \ud\sigma= 
\int_{\mcs}\frac{\chi^2}{\mcr^2}|\Psi_2|\,\ud\sigma\ ,
\ee
where $\ud\sigma = i(P/R)^2\ud\xi\wedge\ud\bar{\xi}$ is the
natural measure on $\mcs$ and $\Psi_2 = - |\Psi_2|$ follows
from \eqref{psi2} and the assumption $m>0$. This shows that
$\chi$ vanishes. Thus, 
\be\label{hnull}
\mch\ \textrm{is null}\Leftrightarrow \lambda=0\ . 
\ee

\section{Null case}\label{nh}
We would like to discuss conditions that $\mcr$ and $P$ should
satisfy for $\mch$ to be a null (nonexpanding) horizon.  In this case
$\lambda=\chi=\nu=0$ (see \eqref{hnull} and \eqref{chinu}) and the
vector tangent to $\mch$ and normal to sections is just $l$ (see
\eqref{tdec}).  Consider the following Newman-Penrose equations
\begin{equation}\begin{aligned}
& \delta\lambda - \bar\delta\mu =
(\rho-\bar\rho)\nu +(\mu-\bar\mu)\pi+\mu(\alpha+\bar\beta)+\\ 
 \label{NPeq3}&+\lambda(\bar\alpha- 3\beta)-\Psi_3 + \Phi_{21}\ ,
 \end{aligned}\end{equation}
 \begin{equation}\begin{aligned}
 & \Delta\lambda -\bar{\delta}\nu = (\bar{\gamma}-3\gamma-\mu-\bar{\mu})
\lambda+\\ \label{NPeq4} &+(3\alpha+\bar{\beta}+\pi-\bar{\tau})\nu-\Psi_4\ .
\end{aligned}\end{equation}
Note that derivatives present in \eqref{NPeq3} and \eqref{NPeq4}
are $\Delta = \nabla_l$, $\delta = \nabla_m$ and 
$\bar{\delta} = \nabla_{\bar{m}}$. These directions
are tangent to $\mch$; this allows us to 
conclude that conditions $\Psi_4=\Psi_3 = 0$ must be
satisfied on $\mch$.
Hence $g$ is of Petrov type D on $\mch$ ($\Psi_2\neq 0$ for
non-vanishing $m$, see
\eqref{psi2} in Appendix A).
From equation (\ref{appbpsi3}) in Appendix A it follows that
on $\mch$
\be\label{psi3}  \frac{3Pm}{R^4}\mcr_{,\bar\xi}+
\frac{P}{2R^2}K_{,\bar\xi} = 0 \ .\ee
This equation can be easily integrated to give $\mcr$
in terms of $K$:
\be\label{RK} \mcr  = 6m(K + a)^{-1}\ ,\ee
where $a=a(u)$ is real. Substituting (\ref{RK}) into  the condition
$\lambda=0$ (see (\ref{appalambda}) in App. A) we get 
\be\label{a1} (P^2K_{,\bar{\xi}})_{,\bar{\xi}}=0\ .\ee 
Note that due to \eqref{RK}, \eqref{a1} is a condition on $P$ only.
Due to the fact that $r$-dependence of the metric is known, further
results hold for arbitrary $r>0$. 
It turns out that  (\ref{a1}) restricts substantially  a set of
solutions of
the RT equation \eqref{RTeq}, regardless of whether the trapped
surfaces are spherical or not:
\begin{tw} \label{J} The only vacuum Robinson-Trautman metrics
 admitting a  non\-ex\-pan\-ding horizon
are the Schwarzschild metric  and the C-metric.
\end{tw}
Proof:  It follows from (\ref{a1}) that 
\be\label{l1} P^2K_{,\bar{\xi}}=f(u,\xi)\ .\ee 
Substituting (\ref{K}) and (\ref{l1}) into (\ref{RTeq}) yields
\be
(fP^{-2})_{,\xi}=3m(P^{-2})_{,u}\ .\label{a2}
\ee
Applying twice the operator $P^2\partial_{\bar{\xi}}+2PP_{,\bar{\xi}}$ 
to (\ref{a2}) we obtain
\be
P^2(K_{,\bar{\xi}})^2=6m(P^2(\ln{P})_{,u\bar{\xi}})_{,\bar{\xi}}\ .\label{a3}
\ee
Conditions (\ref{a1}) and (\ref{a3}) assure that the corresponding
metric is of Petrov type D \cite{KR}. For $f=0$ equations (\ref{l1})
and (\ref{a2}) lead to $P_{,u}=0$ and $K=\const$, hence one obtains
the Schwarzschild metric. According to \cite{KR}, an inspection of
Kinnersley's list of type D vacuum solutions of the Einstein equations
\cite{KINN} shows that for $f\neq 0$ solutions of (\ref{l1}) and
(\ref{a2}) necessarily yield the C-metrics. We present an independent
proof of this property in Appendix B.

In the case of the Schwarzschild metric the trapped surfaces form the
past event horizon $r=2m$. In the case of  the C-metric, by virtue of
(\ref{RK}), equation (\ref{mts}) takes the form
\be\label{ll5} P^2(\ln(K+a))_{,\xi\bar{\xi}} +
\frac{K}{3}-\frac{a}{6} = 0\ .\ee
Substituting (\ref{a21}) to \eqref{ll5} and integrating yields
\be
6mK_{,x}=-\frac{1}{3}K^3-\left(\frac{a^2}{3}+a_1\right)K+aa_1\label{a4}
\ee
where $a_1$ is another function of $u$. Equation (\ref{a4}) coincides
with (\ref{a22}) if $a$ and $a_1$ are constants,
$a_1=b-\frac{a^2}{3}$ and
\be
\frac{a^3}{3}-ab+c=0\ .\label{a5}
\ee
Thus, the C-metric admits locally a nonexpanding horizon. It is known
that this horizon does not admit regular spherical sections (see
e.g. \cite{GKP}) because of conical singularity on each section $\mcs$.
$\Box$

\section{Properties of the surface $r=2m$}\label{eh}
It is interesting to examine the relation between $\mcs$ and the
surface $r=2m$ which plays the role of past event horizon in the
Schwarzschild case. 
In this section  we assume that the surfaces of
constant $r$ and $u$ are diffeomorphic to $S_2$ and the
function $\hP = P/P_S$  is regular on these surfaces.
\begin{tw} \label{r2m} Let $\mcs$  be a regular spheroidal
marginally trapped surface given by \eqref{ciecie} and \eqref{mts}.
Then $\mcs$ crosses the surface $r= 2m$. \end{tw}
Proof: Let 
\be\label{R2md}
\mcr = \frac{2m}{1-h}
\ee
for some smooth function $h<1$. Rewriting \eqref{mts} in
 terms of $h$ we get
\be\label{R2md2}
-2 P^2 [\ln(1-h)]_{,\xi\bar{\xi}} = K-1 + h\  .
\ee
From the Gauss-Bonnet theorem and the regularity of $h$ 
it follows that
\be\int\limits_{\{r=1,\ u=\const\}}^{}\!\!\!\!\!\!
\!\!\!\!\! h =0\ ,\ee
hence $h^{-1}(\{0\}) \neq \emptyset$ which finishes the proof.
$\Box$

Another property of $r=2m$ is the signature of its induced metric for
finite $u$. Obviously, it is null in the Schwarzschild space-time.
The converse statement is also true:
\begin{tw}\label{rtnull} Let $g$ be a vacuum RT metric \eqref{RTmetric}.
 If the hypersurface $r=2m$ is null,
 then $g$ is the Schwarzschild metric.\end{tw}
Proof: Assume that the surface $r=2m$ is null, then in
\eqref{RTmetric} one has to set $H|_{r=2m}=0$ which is equivalent to
\be\label{rtnull1}
K-1 = 4m(\ln P)_{,u}\ .
\ee
Combining \eqref{rtnull1} with the RT  equation
\eqref{RTeq} we get
\be\label{rtnull2}
\partial_u\left(
4m(\ln P)_{,\xi\bar{\xi}}-3mP^{-2}\right)=0\ .
\ee
Hence
\be\label{rtnull3}
(\ln P)_{,\xi\bar{\xi}}-\frac{3}{4}P^{-2}=f(\xi,\bar{\xi})
\ee
and \be\label{rtnull3.5}
K = \frac{3}{2} + 2P^2 f\ee 
follows. 
Differentiating  $K$ with respect to $u$ and using \eqref{rtnull1}
and \eqref{rtnull3.5} we obtain
\be\label{rtnull4}
K_{,u} = 4PP_{,u} f = 2(\ln P)_{,u} 2P^2 f = \frac{1}{2m}(K-1)(K-3/2)\ .
\ee
This equation can be integrated to give
\be\label{rtnull5}
\frac{K-1}{K-3/2}=h(\xi,\bar{\xi})e^{-u/4m}
\ee
for some $u$-independent function $h$. Now from
\eqref{rtnull3.5} and \eqref{rtnull5} we have
\be\label{hreg}
K - 1 = h e^{-u/4m} 2P^2 f\ .
\ee
 Solving \eqref{rtnull5} 
for $K$ and using  \eqref{rtnull3.5} we have
\be\label{rtnull6}
P^2 = \frac{1}{4f}\frac{1}{he^{-u/4m}-1}\ .
\ee
Inserting \eqref{rtnull6} into the RT equation yields a
polynomial of degree 4 in $U= \exp(-u/4m)$ equal to zero:
\be\label{rtnullp}
(h_{,\xi\bar{\xi}} - 6 f h) U
- (h h_{,\xi\bar{\xi}} - 2 |h_{,\xi}|^2 - 18 fh^2)U^2
- 18f(hU)^3 +6f(hU)^4=0\ .
\ee
Since $h$ is $u$-independent, it has to vanish
identically, and from \eqref{hreg} we get $K=1$
and $P = P_S$. Hence $g$ is the Schwarzschild metric.$\Box$

\section*{Acknowledgments}
The authors would like to thank P. Chru{\'s}ciel and M. Korzy{\'n}ski
for helpful comments.

This work was  supported by the Polish Ministry
of Science and Higher Education grant 1 PO3B 075 29.

\appendix{Spin coefficients and Weyl scalars
 for tetrad (\ref{tetrad})}
We follow the notation from \cite{KR} do derive the 
spin coefficients and Weyl scalars of \eqref{tetrad}.
\be
\alpha  = \frac{P_{,\bar{\xi}}}{2r}-\frac{P\mcr_{,\bar{\xi}}}{r^2}\ ,\ 
\ee
\be
\beta = -\frac{P_{,\xi}}{2r}\ ,\ 
\ee
\be
\tau = \bar{\pi} = -\frac{P}{r^2}\mcr_{,\xi}\ ,
\ee
\be\label{appalambda}
\lambda  = (\partial_{\bar{\xi}}+\mcr_{,\bar{\xi}}\partial_r)
\left(\frac{P^2}{r^2}\mcr_{,\bar{\xi}}\right)\ ,\ 
\ee
\be
\rho=-\frac{1}{r}\ ,\ 
\ee
\be
\gamma  = \frac{1}{2r^3}[- 2 P^2|\mcr_{,\xi}|^2 + r^3 H_{,r} +
rP(P_{,\bar{\xi}}\mcr_{,\xi}-\cc)]\ ,
\ee
\be\label{aanu}
\nu  = \frac{P}{r}(\partial_{\bar{\xi}}+\mcr_{,\bar{\xi}}\partial_r)
\left(\mcr_{,u} + H + \frac{P^2}{r^2}
|\mcr_{,\xi}|^2\right)\ ,
\ee
\be
\mu  = -\frac{H}{r}-(\ln P)_{,u}+\frac{P^2}{r^2}\mcr_{,\xi\bar{\xi}}
-\frac{P^2}{r^3}|\mcr_{,\xi}|^2\ ,
\ee
\be\label{psi2}
\Psi_2= -\frac{m}{r^3}\ ,\ee
\be\label{appbpsi3}
\Psi_3 = \frac{3Pm}{r^4}\mcr_{,\bar{\xi}} + \frac{P}{2r^2}K_{,\bar{\xi}}\ .\ee
Note that from the definition of $\chi$ in \eqref{tdec} and \eqref{aanu}
 it follows that
\be\label{chinu}
\nu = \nabla_{\bm}\chi\ .
\ee

\appendix{Robinson-Trautman metrics of type D}\label{AppC}
In section 4 we have shown that a necessary condition for $\mch$ to be
null is that the Robinson-Trautman metric is of type D. In this appendix we
will explicitly solve equations (\ref{l1}) and (\ref{a2}) equivalent
to the condition $\lambda=0$ and the Robinson-Trautman equation. Since
$f=0$ corresponds to the Schwarzschild metric or a flat one we will
assume here that $f\neq 0$.

Let
\be 
\xi'=h(u,\xi)\label{B1}
\ee
where
\be
h_{,\xi}=f^{-1}\label{B2}
\ee
(note that transformation (\ref{B1}) does not preserve the
Robinson-Trautman equation (\ref{RTeq})). In terms of coordinates
$\xi',\ \bar{\xi}'$ and $u'=u$ equation (\ref{l1}) reads
\be\label{B3} P^2K_{,\bar{\xi}'}=|f|^2\ .\ee 
Hence, $K=K(u',x')$, where $x'=\Re\xi'$ and
\be
P^2=\frac{2}{K_{,x'}|h_{,\xi}|^2}\ .\label{B4}
\ee
Substituting (\ref{B4}) into definition (\ref{K}) of $K$ yields
\be
(\ln{K_{x'}})_{,x'x'}=-2KK_{,x'}\ .\label{B5}
\ee
Integrating twice equation (\ref{B5}) we get
\be
K_{,x'}=-\frac{1}{3}K^3+bK+c\ ,\label{B6}
\ee
where $b$ and $c$ are functions of $u$.

It follows from (\ref{B4}) that 
\be
(\ln{P})_{,u}=-\frac{1}{2}(\ln{K_{,x'}})_{,u'}-
\frac{1}{2}(\ln{K_{,x'}})_{,x'}\Re h_{,u}-
(\ln{|h_{,\xi}|})_{,u}\label{a8}
\ee
and the Robinson-Trautman equation takes the form
\be
(\ln{K_{,x'}})_{,u'}+(\ln{K_{,x'}})_{,x'}\left(\Re{h}_{,u}-
\frac{1}{6m}\right)+2(\ln{|h_{,\xi}|})_{,u}=0\ .\label{a9}
\ee
By virtue of (\ref{B6}) acting on (\ref{a9}) with the operator
$\partial_{\xi'}\partial_{\bar\xi'}$=$|h_{,\xi}|^{-2}
\partial_{\xi}\partial_{\bar\xi}$
yields
\be
(\ln{(KK_{,x'})})_{,u'}+(\ln{(KK_{,x'})})_{,x'}\left(\Re{h}_{,u}
-\frac{1}{6m}\right)+2(\ln{|h_{,\xi}|})_{,u}=0\label{a10}
\ee
Subtracting (\ref{a9}) from (\ref{a10}) implies
\be
\frac{K_{,u'}}{K_{,x'}}+\Re{h}_{,u}-\frac{1}{6m}=0\ .\label{a11}
\ee
Applying  $\partial_{\xi'}\partial_{\bar\xi'}$ 
to (\ref{a11}) shows that $K_{,u'}/K_{,x'}$ is linear in $x'$ and 
\be
h_{,u'}=c_1(u)h+c_2(u)\label{a12}
\ee
where $c_1$ and $c_2$ are, respectively, a real and a 
complex function of $u$. Hence 
\be
h=\frac{2}{c_3(u)}h'(\xi)+c_4(u)\label{a13}
\ee
where $c_3$ is a real function of $u$ and $h'$ is a function of
$\xi$. Note that transformation $\xi\rightarrow h'(\xi)$, accompanied
by $P\rightarrow P|h'_{,\xi}|$ preserves the Robinson-Trautman
metric. Thus, locally, without loss of
generality we can assume $h'(\xi)=\xi$.  
In this case substituting (\ref{a13}) into (\ref{B4}) yields
\be
P^2=\frac{c_3}{K_{,x}}\label{a14}
\ee
where now $K=K(u,x)$ is considered as a function of $u$ and
$x=\Re{\xi}$.  Given (\ref{a14}) we see that $P=P(u,x)$ and equation
(\ref{a2}) takes the form
\be
c_3P_{,x}-12mP_{,u}=0\label{a16}
\ee
It follows from (\ref{a16}) that $P$ 
is a function of the variable $s=x+d(u)$, where 
\be
d_{,u}=\frac{c_3}{12m}\ .\label{a16a}
\ee
 Equation (\ref{a14}) shows that 
\be
K=c_3(K'(s)+c_5(u))\label{a17}
\ee
\be
P^2=\frac{1}{K'_{,s}}\ ,\label{a18}
\ee
where $K'$ and $c_5$ are functions of $s$ and $u$, respectively.
Substituting (\ref{a17}) and (\ref{a18}) into relation (\ref{K})
implies that $c_3$ and $c_5$ are constants. Due to (\ref{a16a}) and
the remaining freedom of linear transformations of $\xi$ we can
assume that
\be
K=K(x+u)\ ,\ \ P^2=\frac{12m}{K_{,x}}\ .\label{a21}
\ee
Now equation (\ref{B6}) takes the form
\be
6mK_{,x}=-\frac{1}{3}K^3+bK+c\label{a22}
\ee
where $b$ and $c$ are constants.
Robinson-Traut\-man metrics
satisfying conditions (\ref{a21}) and (\ref{a22}) are equivalent to
the C-metrics \cite{KIN}. Thus, the Schwarz\-schild metric and the
C-metric are the only vacuum Robinson-Traut\-man metrics which satisfy
condition (\ref{a1}).


\begin{thebibliography}{50}

\bibitem{RT} Robinson I., Trautman A.,
{\it  Some spherical gravitational waves in general relativity},
Proc. Roy. Soc. A {\bf 265} 463 (1962)

\bibitem{KR} Stephani H, Kramer D, MacCallum M A H,
Hoenselaers C and Herlt E
{\it Exact Solutions to Einstein's Field Equations, Second Edition},
(Cambridge: Cambridge University Press) (2003)

\bibitem{CHR} Chru\'sciel P. T.,
{\it  On the global structure of Robinson--Trautman space-times},
 Proc. R. Soc. Lond. A {\bf 436} 299 (1992) 

\bibitem{AK} Ashtekar A.,  Krishnan B.,
 {\it Isolated and Dynamical Horizons and Their Applications},
Living Rev. Relativity {\bf 7}, (2004), 10 [Online Article]:
cited [\today], http://www.livingreviews.org/lrr-2004-10

\bibitem{TOD} Tod P., 
{\it Analogues of the past horizon in Robinson-Trautman space-times},
Class. Quantum Grav. {\bf 8} 1159 (1989)

\bibitem{CL} Chow E. W. M., Lun A. W.-C.,
 {\it Apparent Horizons in Vacuum Robinson-Trautman Spacetimes},
J. Austral. Math. Soc. Ser. B {\bf 41} 217 (1999), 
http://arxiv.org/abs/gr-qc/9503065 (1995)

\bibitem{GKP} Griffiths J. B., 
Krtou{\u s} P., and Podolsk{\' y} J.,
 {\it Interpreting the C-metric,}
Class. Quantum Grav. {\bf 23}  6745 (2006)

\bibitem{AMS} Andersson L., Mars M., Simon W.,
{\it Local Existence of Dynamical and Trapping Horizons},
Phys. Rev. Lett. {\bf 95}, 111102 (2005)
 

\bibitem{MK} Korzy\'nski M., 
 {\it Isolated and dynamical horizons from a common perspective}, 
 Phys. Rev. {\bf D74}, 104029 (2006) 
 
\bibitem{KINN} Kinnersley W.,
{\it Type D vacuum metrics},
J. Math. Phys. {\bf 10}, 1195 (1969)

\bibitem{KIN} Kinnersley W.,
{\it  Field of an Arbitrarily Accelerating Point Mass,}
Phys. Rev. {\bf 186}, 1335 (1969)

\end{thebibliography}
\end{document}